# ALMA Cycle 0 Publication Statistics

Felix Stoehr[1]
Uta Grothkopf[1]
Silvia Meakins[1]
Marsha Bishop[2]
Ayako Uchida[3]
Leonardo Testi[1]
Daisuke Iono[3]
Kenichi Tatematsu[3]
Al Wootten[2]

[1] ESO
[2] National Radio Astronomy Observatory, Charlottesville, USA
[3] National Astronomical Observatory of Japan, Mitaka, Tokyo, Japan

The scientific impact of a facility is the most important measure of its success. Monitoring and analysing the scientific return can help to modify and optimise operations and adapt to the changing needs of scientific research. The methodology that we have developed to monitor the scientific productivity of the ALMA Observatory, as well as the first results, are described. We focus on the outcome of the first cycle (Cycle 0) of ALMA Early Science operations. Despite the fact that only two years have passed since the completion of Cycle 0 and operations have already changed substantially, this analysis confirms the effectiveness of the underlying concepts. We find that ALMA is fulfilling its promise as a transformational facility for the observation of the Universe in the submillimetre.

## Introduction

The Atacama Large Millimeter/submillimeter Array (ALMA) was designed and built as a world-leading submillimetre observatory that would produce transformational science on a broad range of astrophysical topics. Early Science observations began in 2011 in response to increasing pressure from the astronomical community to use the facility, even though only a small fraction of the final capabilities were available. Eventually, ALMA was offered in Cycle 0 for observations with sixteen 12-metre antennas, four frequency bands covering 84–116 GHz (Band 3), 211–275 GHz (Band 6), 275–373 GHz (Band 7), and 602–720 GHz (Band 9), baseline lengths of up to 400 metres and limited observing modes and correlator capabilities. The capabilities that were offered correspond to about 25 % of the final collecting area, 40 % of the frequency bands, only 2 % of the achievable angular resolution and a small fraction of the observing and correlator modes of the completed observatory.

Science observations for ALMA Cycle 0 were carried out from 1 October 2011 through 31 January 2013. The vast majority of the calibrated datasets were delivered to the users by the first quarter of 2013; the last dataset was delivered in mid-August 2013.

## Methodology

In this section, we summarise the methodology applied in order to compile the ALMA bibliography. A more detailed version, along with lessons learned from the first two years of maintaining this important resource, can be found in Meakins et al. (2014).

The ALMA bibliographic database is jointly maintained by the librarians at ESO and the National Radio Astronomy Observatory of the USA (NRAO), together with administrative support from the National Astronomical Observatory of Japan (NAOJ). The bibliographic database has been built using the extensive experience accumulated by the librarians in the context of similar telescope bibliographies at their institutes (NRAOpapers[1] and ESO telbib[2]).

For the purposes of this study, only refereed publications that appeared in print at the time of writing and that make direct use of ALMA data are considered. Papers that only cite values or results derived from ALMA data from the literature are excluded. Also excluded are publications that describe instrumentation or software, merely mention ongoing projects, suggest future observations, or develop models or run simulations without actual use of data. This definition corresponds to the *Best Practices for Creating a Telescope Bibliography*, the IAU/Commission 5 endorsed guidelines for the construction and maintenance of bibliographies related to observatory or telescope bibliographies[3].

Compilation of the ALMA bibliography consists of a multi-step workflow that starts with the screening of core journals in astronomy as well as of various other journals and arXiv/astro-ph eprints (journals routinely screened are A&A, A&Arv, AJ, AN, ApJ, ApJS, ARA&A, EM&P, Icarus, MNRAS, Nature, NewA, NewAR, PASJ, PASP, P&SS, and Science). However, in this article only papers that have been fully published and have received their final bibcode are considered. All papers that contain at least one ALMA-related keyword are visually inspected by the librarians to decide whether or not they qualify for inclusion in the database. Bibliographic details (authors, titles, publication details, etc.) along with other metadata are imported from the Smithsonian Astrophysical Observatory (SAO)/National Aeronautics and Space Administration (NASA) Astronomical Data System (ADS).

For each record in the bibliography, a variety of parameters are identified and added by the librarians. Most importantly, this includes the ALMA project code(s) of the observations that were used, the ALMA partner to whose share the observing time was attributed and the observing type (Standard, Large, Target of Opportunity, Science Verification, etc.). At this point, archive flags are assigned to project codes without overlap between authors of the paper and Principal Investigators/Co-Investigators (PIs/CoIs) of the Observing Programme (see section on archival research below).

ALMA's User Policies require authors not only to include the official ALMA acknowledgement text in their publications, but also the ID number of the ALMA Programme of the data used, formatted according to the ADS data-tag[4] specification. This allows the creation of links between publications and data, which forms not only the basis of most of the results presented here, but also allows readers of the articles to immediately access the data in the ALMA Science Archive[5]. Additionally, in the near future, archival researchers can find publications related to the data records in the ALMA Science Archive.

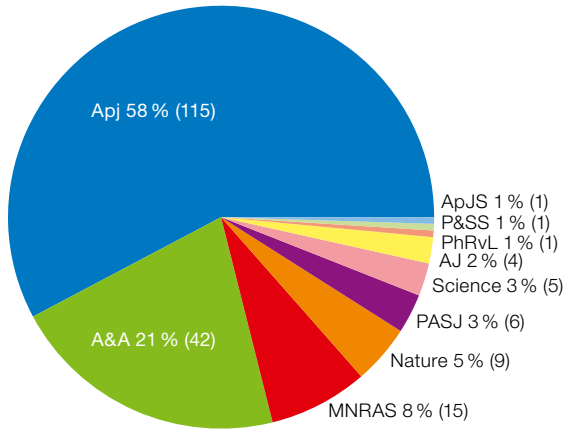

Figure 1. Distribution of ALMA Cycle 0 publications by refereed journal.

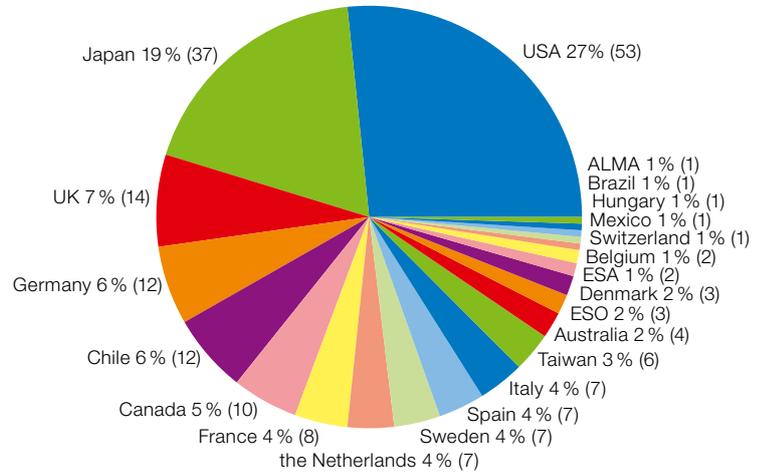

Figure 2. Geographical distribution of the first authors of ALMA Cycle 0 publications.

A large and increasing fraction of authors (about 85 %) know these policies well and place the statement and the data tag in their publications. Furthermore, we regularly check the arXiv.org preprint server and contact the authors by email if the acknowledgement or data tag are missing from a paper or are incorrect. Feedback from contacted authors is extremely positive with the result that 98 % of the ALMA publications that go to press carry the acknowledgement statement and data tag. The remaining publications without data tags are those that were put on the preprint server only when the publication was already in press. In those cases a member of the library staff identifies the related ALMA Programmes manually and adds the information to the publication database.

Experience from two years of maintenance of the ALMA bibliography has shown clearly that this personal communication is essential to: (a) raise the awareness of authors regarding the ALMA policy concerning the use of ALMA data in the acknowledgements; and (b) to establish a bibliography that is as complete and correct as possible.

## General overview of the publications

At the time of writing the number of refereed publications using ALMA data has reached a total of 308 papers, mostly using Cycle 0 data (199, or 65 %) and Science Verification (SV) observations

(76, or 25 %). There are 47 (15 %) papers based on Cycle 1 and 2 data including 3, or 1 %, based on Cycle 1 and 2 Director's Discretionary Time (DDT). In Cycle 0 DDT was not offered. As a single paper can make use of several ALMA Programmes, the sum over the values of the different cycles is larger than 100 %.

In the following we concentrate on publications from Cycle 0 data. All Cycle 0 data were delivered more than two years ago, which allows us to carry out a global and unbiased first analysis.

## Journals

A significant fraction of the Cycle 0 ALMA publications (7 %) appeared in non-sector specific, high-impact journals (*Nature* and *Science*), demonstrating the transformational nature of the ALMA Observatory, even in the limited configuration available for Cycle 0 observations. Such a high fraction is uncommon for ground-based observatories in the first years of operations and more in line with that of space missions. The premier journals for ALMA publications are *The Astrophysical Journal* (ApJ) and *Astronomy & Astrophysics*, which together account for 79 % of the refereed publications (see Figure 1). It is noteworthy that nearly a quarter of all papers are published as Letters — in addition to the publications in *Nature* and *Science*. For all regions, ApJ publishes most of the Letters. This, together with the fact that most North American (NA) authors and many East Asia (EA) authors publish in ApJ, explains the large fraction of ApJ publications.

## Countries

The overall publication numbers (Figure 2) are well balanced among the three ALMA partners (NA, EA and ESO [EU]) and are in line, within the uncertainties, with their shares in the Observatory (37.5 % each for ESO (EU) and NA, 25 % for EA). There is a marginally larger fraction of first authors from the ESO members (EU 37 %, NA 33 %, EA 20 %, Chile 4 %, other 6 %), but this is not statistically significant. The fraction of first authors who use data obtained through observing time attributed to a different partner from that of their home institute hovers around 15 %, indicating a healthy mix of the partners in the research groups and mobility of the researchers across the regions.

## Distribution by science area

The right-hand side of Figure 3 shows the distribution of Cycle 0 publications according to their scientific categories. This distribution is relatively well balanced. Notable differences in the distribution of scientific categories of accepted proposals that received ALMA data (Figure 3, left) are only observed in the galaxy evolution and Solar System categories. A single project is responsible for the over-proportionally large number of publications in the galaxy evolution category. Solar System data in Cycle 0 turned out to be very difficult to obtain, calibrate and analyse and this is considered as the reason for the lower number of publications so far.

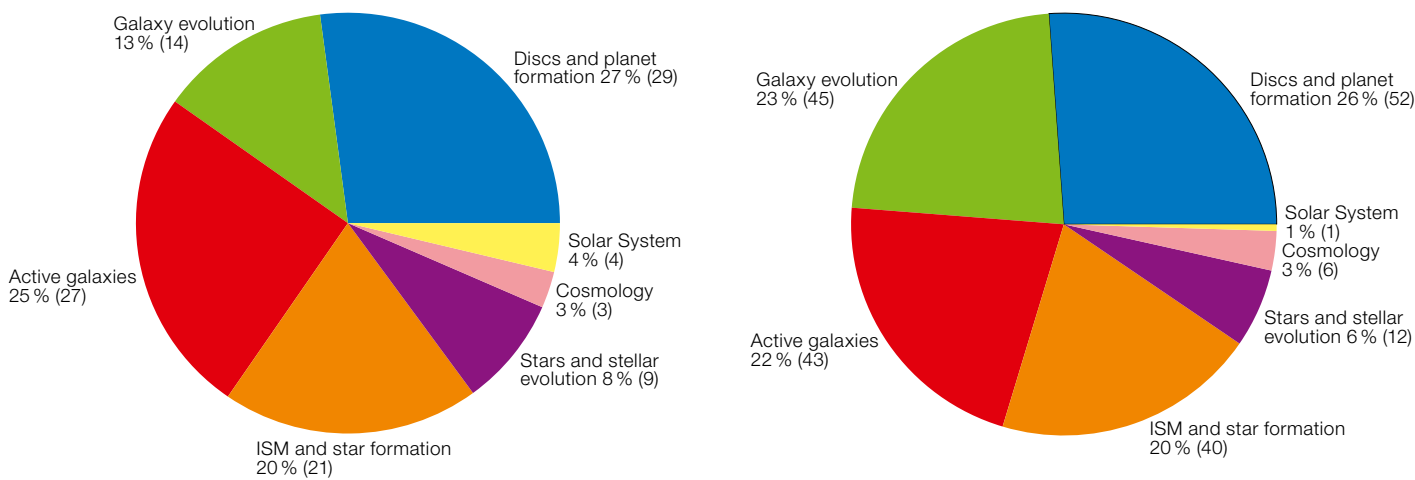

**Figure 3.** Distribution of the scientific categories of allocated Cycle 0 high priority Programmes (left) and the resulting publications (right).

### Publication fraction

An important measure for a facility is the fraction of Programmes that did receive data and that result in a publication (PI or archival). In other words: how effectively can the data be converted into published science? Figure 4 shows this fraction as a function of time since the data became available to the PI. For each bin only the Programmes that are available for that timespan are taken into account, hence the plot is not strictly cumulative for the left-most bins, which suffer from low number statistics. However we can conclude that the ALMA publication fraction for Cycle 0 is roughly 85 %, an extremely high value even when compared to space programmes. Although fully comparable numbers are not available, facilities like the Very Large Telescope (VLT), Hubble Space Telescope (HST) or the X-ray Multi-Mirror Mission (XMM-Newton) have publication fractions in the range of 50–75 % (Sterzik et al., p. 2, private communication and Ness et al. [2013] respectively).

**Figure 4.** The fraction of programmes that have had at least one related publication as a function of the time since the data was made available to the PI. The light blue lines show 1σ standard deviations assuming 50 % publication probability and weighted by the number of projects that have not yet reached the age to contribute to a bin.

The high publication fraction not only indicates that ALMA is a transformational facility, but also suggests that the quality of the ALMA data is extremely high and that the effort of providing science-grade data to users allowed nearly all Cycle 0 PIs, including those with limited previous interferometric or millimetre-wave background, to make effective scientific use of the data. This conclusion is also supported by the results of the ALMA User Surveys.

We speculate that the exceptionally high publication fraction suggests that the Cycle 0 ALMA Proposal Review Committee (APRC) awarded time preferably to high science profile Programmes with a high probability of success (low-risk approach in the selection of the science programme); possibly as a consequence also of the extremely high oversubscription rate of nine for Cycle 0.

**Figure 5.** The distribution of the time spans from the availability of the data for a Programme until the first resulting publication.

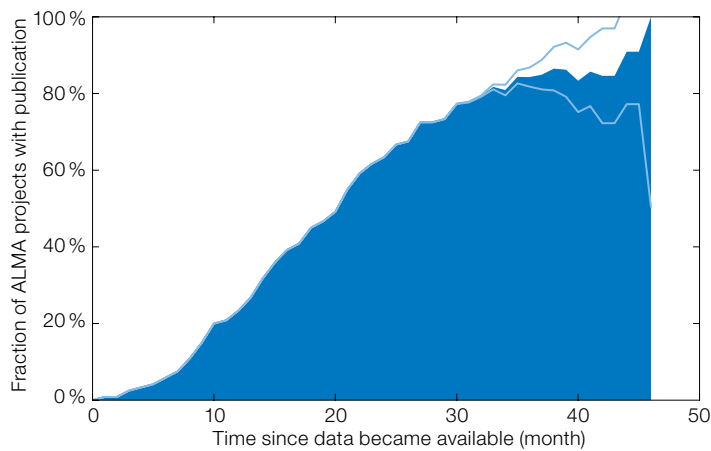

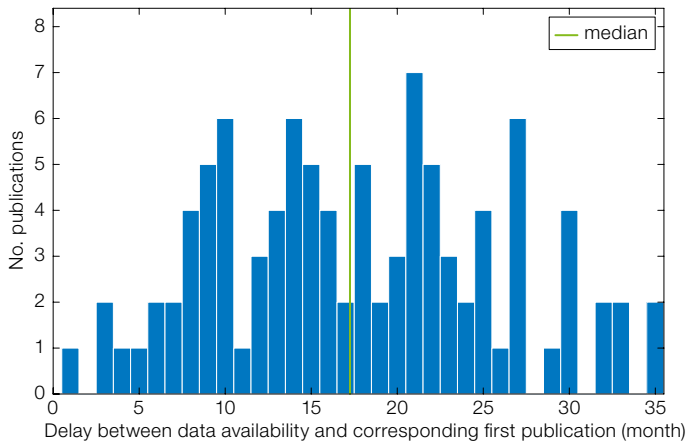

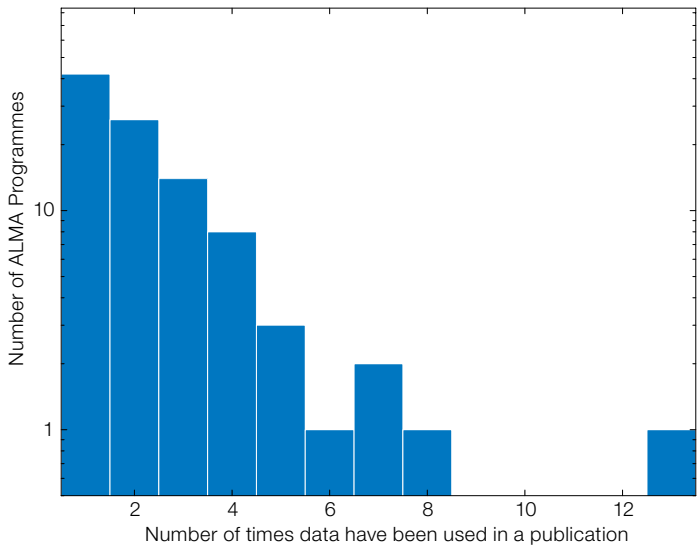

Figure 6. Distribution of the number of times data from a given Programme has been used in a refereed publication.

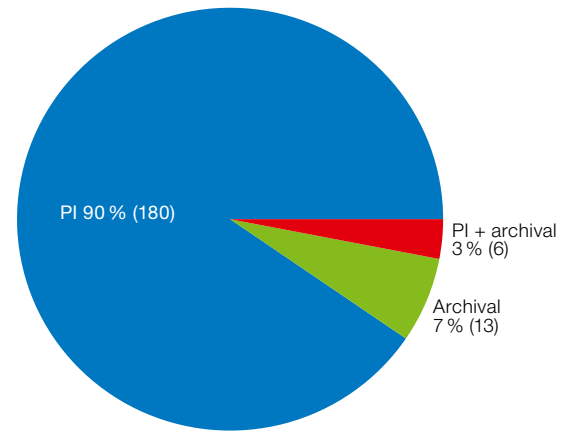

Figure 7. Distribution of publications that use only PI data, only archival data, or a combination of both.

### Publication delay

Figure 5 shows the time that it takes from the delivery of the data to the PI to the first publication using the data, either by the original proposers or by an archival research team. Up to 32 months the statistics are complete and include 80 % of all Programmes. We can therefore safely conclude that the large majority of the ALMA Cycle 0 data were published extremely rapidly. The median value is 1.4 years, which is considerably faster than standard programmes of space missions like XMM or Chandra (both about 2.3 years: Rots et al., 2012; Ness et al., 2013). The ALMA Cycle 0 publication delay is even comparable to the delay for XMM Target of Opportunity Programmes, which have a six-month proprietary period. We believe that the high pressure to obtain time, the transformational nature of ALMA and the fact that ALMA is the first submillimetre observatory in the southern hemisphere, have all contributed to the fast publication rate of Cycle 0 Programmes.

In addition to the novelty of ALMA, we also attribute this effect to the fact that users did not need to spend much time on data reduction, as they received fully calibrated data and reference images directly from ALMA. In addition, the fact that ALMA provides a comprehensive user-support model, which offers Helpdesk support on three continents as well as face-to-face support from proposal preparation to data reduction and analysis (see Hatziminaoglou et al., p. 24), has a very positive impact on the usability of ALMA data by the community. Both claims are again well supported by the results of the ALMA User Surveys.

### Multiple use of the data

ALMA Cycle 0 Programmes, which have resulted in at least one publication, produced on average about two publications per Programme. Experience from other observatories shows that the ALMA value is expected to rise with time. After four years, this value for ALMA is already identical to a similar quantity for the VLT averaged over nine years (Sterzik et al., p. 2). More than half of all published ALMA Programmes were used in at least two publications (Figure 6), with Programme 2011.0.00294.S *More than LESS: The first fully-identified submillimetre survey* (PI: Ian Smail) being by far the most published Cycle 0 Programme, with 13 publications to date.

These numbers confirm the richness and quality of ALMA data and indicate that the ALMA archive is a very valuable asset to maximise the scientific return of the facility.

### Archival research

A paper is considered as using archival data in the case where there is no overlap between the authors and PI & CoIs of the original Programme from which data were used. Where a publication makes use of archival as well as proprietary data, the paper is tagged "PI + archival".

The fraction of publications that made use of ALMA archival data has been steadily increasing and has now reached 7 % for purely archival publications and an additional 3 % for publications making use of Cycle 0 archival data together with ALMA PI data (Figure 7). We expect this fraction to continue to grow with time; even more so for future cycles once full-cube science-grade pipeline products are available. Note that although SV data are also tagged "archival", in the statistics presented above, only publications making use of data taken as part of Cycle 0 Programmes have been taken into account.

### Publications from SV data

In order to test new capabilities, ALMA has been regularly executing SV projects[6]. The resulting data were reduced and made publicly available without a proprietary period. These data have, so far, led to a very large number of refereed publications. The 17 SV projects have been used in 76 publications (4.5 publications per project). Due to the highly competitive setting, without a proprietary period, the median timespan between the release of the data and the first corresponding publication is only four months. Multi-use of the data has also been very high, with one SV dataset (2011.0.00009. SV, Orion KL Band 6) having been used in 18 publications.

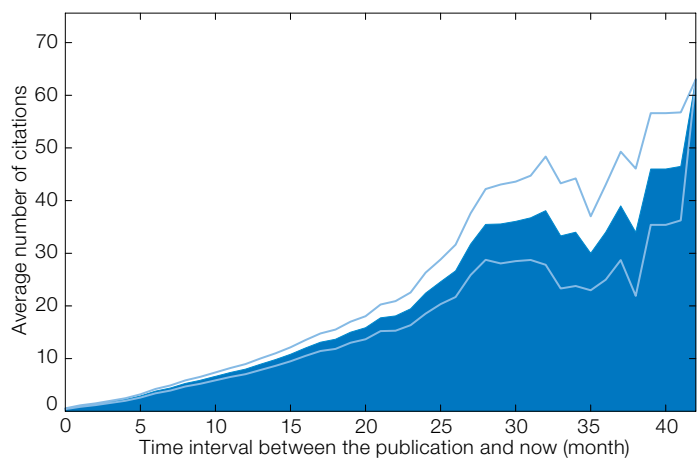

Figure 8. Average number of citations of ALMA publications as a function of time from publication. The light blue lines show the Poisson uncertainty for the counts in each time bin.

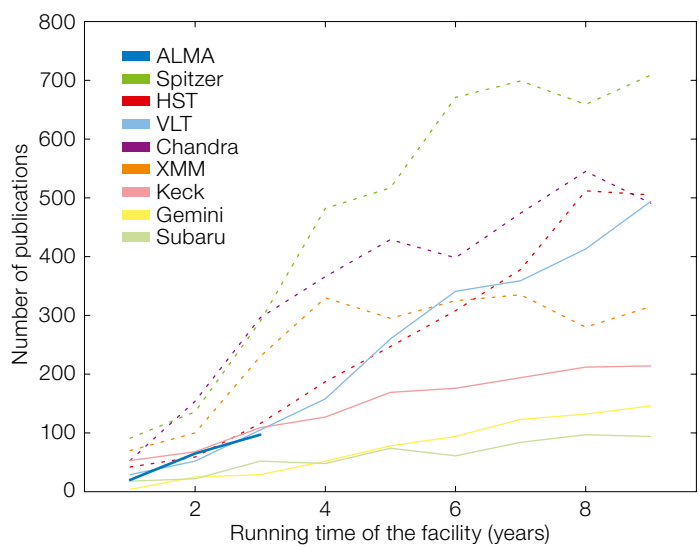

Figure 9. Comparison of the evolution of the number of publications since the first publication for some of the large space-based (dotted) and ground-based (solid, including ALMA as a thick blue line) astronomical facilities.

## Scientific impact

One proxy measurement for scientific impact is the frequency at which publications are cited in the scientific literature. We track citations to ALMA papers using the SAO/NASA ADS services. It is obviously too early to draw final conclusions at this stage; the numbers reported should be considered only as trends. Figure 8 shows the evolution of the average number of citations to Cycle 0 ALMA papers.

We find that, on average, each Cycle 0 publication receives about ten citations per year, which is higher than the 6.5 citations per year measured for ALMA SV publications. Comparing the average number of citations per paper per year to other facilities has to be done with great care because: (a) the value changes with time; and (b) the methodologies and policies for inclusion and exclusion may vary among observatories. The average number of citations per paper per year for VLT publications is four (measured over the last 15 years and computed from telbib[2] using citation numbers from SAO/NASA

ADS), which is identical to the value for XMM (measured over 13 years, Ness et al., 2013). When taking only the last three years of VLT publications into account to match the period with the ALMA data, then the average number of citations per paper per year is about five.

When comparing ALMA data papers to those published in 2012–2014, we find that there are one, six and three ALMA papers, respectively, in the most cited 1% of all refereed astronomy articles in 2012, 2013 and 2014. Thus about 4% of all ALMA publications make it into the top 1% most-cited papers per year (computed from SAO/NASA ADS).

## Comparison with other facilities

In Figure 9 we compare the number of publications making use of data from major astronomical facilities as a function of the time that they have been operational. In order to be comparable, publications from all ALMA cycles as well as publications using SV data have been included.

Space observatories (dotted lines in Figure 9) have traditionally had much steeper rises in publication numbers than ground-based facilities. This effect may be due to the fact that space probes need to be fully prepared at launch and can ramp up very quickly to full operation, whereas ground-based facilities are usually commissioned gradually while some operations have already begun. In the early years, ALMA's scientific output — measured here by the number of scientific papers — is in line with that of HST, VLT or Keck.

## Outlook

In order to close the loop with the PIs and to learn the reasons that prevented PIs from publishing their data, ALMA has initiated a survey, which is sent to PIs of Programmes without a publication two years after the end of the proprietary period. The results of this survey will become part of the monitored elements to constantly improve the operational model and thus optimise the scientific return of ALMA.

## Conclusions

We have analysed the publications of the first Early Science proposal cycle (Cycle 0) of ALMA. Although only two years have passed since the end of this cycle, and more publications making use of Cycle 0 data can be expected, a number of conclusions can already be reached.

ALMA is indeed the transformational facility it was designed to be and has effectively opened the submillimetre wavelength regime to general user observations, with Bands 7 and 9 (275–950 GHz) the basis of 64 % of all Cycle 0 publications. All originally foreseen scientific categories — with the exception of observations of the Sun which are not yet fully commissioned — are represented. We find that a very high fraction of Programmes (roughly 85 %) resulted in one or more publications, and on aver-age, each Cycle 0 publication receives ten citations per year. Publications appear very fast once the data becomes available to the PIs (1.4 years) and the fraction of publications making use of archival data is rising (currently at ~ 10 %). The evolution of the total number of publications is similar to the early evolution of HST, VLT and Keck.

Science Verification data has led to a very large number of publications (76 publications from 17 SV datasets) with very short publication delays (four months), large multiple use (maximum 18) and 6.5 citations per publication per year.

We attribute the success of Cycle 0 in terms of scientific impact to the novelty of the instrument, the high quality of the data, the fact that ALMA delivers fully calibrated and quality-controlled data as well as to the extensive user support in three continents.

### Acknowledgements

The ALMA bibliography makes use of the SAO/NASA ADS[7].

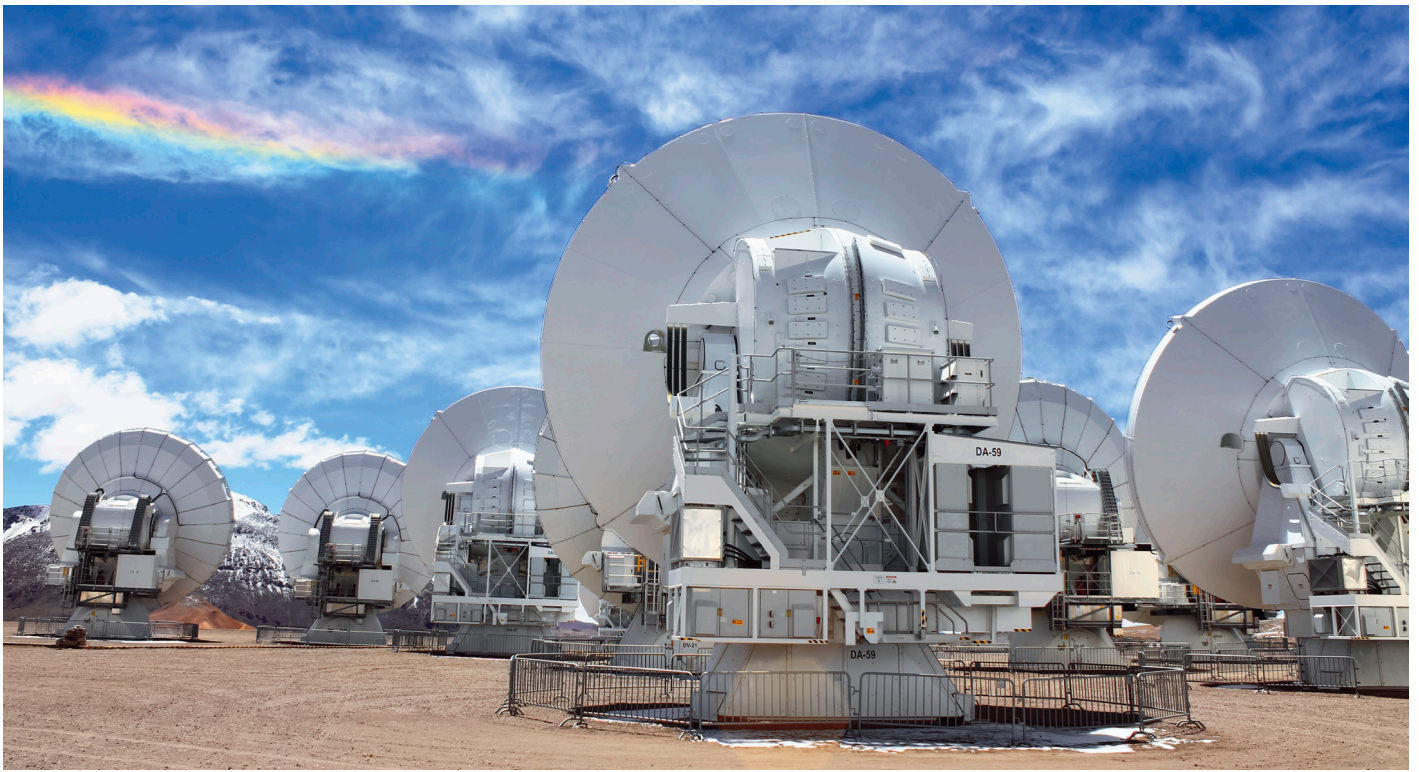

Part of the Atacama Large Millimeter/submillimeter Array (ALMA) during day time. The rare optical phenomenon known as a circumhorizontal arc is formed by the refraction of sunlight in plate-shaped ice crystals suspended in the atmosphere.

J. C. Rojas/ESO